\documentclass[aps,prb,twocolumn,superscriptaddress,showpacs,floatfix]{revtex4-1}

\usepackage{graphicx,latexsym}
\usepackage{amsmath,amssymb,amsfonts}
\usepackage{bm}
\usepackage{braket}
\newcommand{\bmr}{\bm{r}}
\newcommand{\bmq}{\bm{q}}

\begin{document}



\title{Bosonic integer quantum Hall effect as topological pumping}
\author{Masaya Nakagawa}
\email{m.nakagawa@scphys.kyoto-u.ac.jp}
\affiliation{Department of Physics, Kyoto University, Kyoto 606-8502, Japan}
\author{Shunsuke Furukawa}
\affiliation{Department of Physics, University of Tokyo, 7-3-1 Hongo, Bunkyo-ku, Tokyo 113-0033, Japan}
\date{\today}




\begin{abstract}
Based on a quasi-one-dimensional limit of quantum Hall states on a thin torus, we construct a model of interaction-induced topological pumping which mimics the Hall response of the bosonic integer quantum Hall (BIQH) state. The quasi-one-dimensional counterpart of the BIQH state is identified as the Haldane phase composed of two-component bosons which form effective spin-$1$ degrees of freedom. 
An adiabatic change between the Haldane phase and trivial Mott insulators constitutes {\it off-diagonal} topological pumping in which the translation of the lattice potential for one component induces a current in the other. The mechanism of this pumping is interpreted in terms of changes in polarizations between symmetry-protected quantized values. 
\end{abstract}

\pacs{05.30.Jp, 03.75.Mn, 73.43.Cd}


\maketitle


\section{Introduction}

Adiabatic change of parameters of a Hamiltonian sometimes causes nontrivial effects which cannot be found in its instantaneous ground state. Topological pumping, originally proposed by Thouless,\cite{Thouless} provides a prototypical example of such phenomena. Thouless considered a one-dimensional (1D) band insulator in a periodic lattice potential $V(x)=V(x+a)$. Let us consider an adiabatic shift of the lattice potential $V(x-at/T)$ parametrized by $t$. Since the lattice potential is periodic, so is the Hamiltonian with respect to its parameter: $H(t+T)=H(t)$. The one-particle quantum states in an energy band are then specified by the wave number $k_x$ and the parameter $t$. When the lattice potential is adiabatically shifted by varying $t$, 
the particles move with the lattice, and the Hamiltonian returns to its original form after a shift by one lattice spacing. The ground state also returns to the initial one as long as the particles stay in the same band. 
However, the change in the ground state during the cycle can cause a nonvanishing particle current. The total current over
one cycle is given by the Chern number, which takes an integer and characterizes topologically distinct classes of the set of one-particle states defined in the $(k_x, t)$ plane. Hence the total current is quantized, and this phenomenon is called the topological Thouless pumping. If we identify the $(k_x,t)$ plane with the two-dimensional (2D) reciprocal space, we find that the Thouless pumping shares the same origin as the integer quantum Hall (QH) effect,\cite{TKNN,Kohmoto} in which the quantized pumping corresponds to the quantized Hall conductivity.
Owing to its topological nature, the topological pumping is robust against small deformation of the pumping protocol, and can be 
realized by various types of cycles using lattice models.\cite{RiceMele, NiuReview, Shindou, BergLevinAltman} After almost 30 years since Thouless's prediction, the topological pumping was finally realized experimentally by using ultracold atoms in optical superlattices,\cite{Nakajima, Lohse, Schweizer} in which the periodic change of the lattice potential was created by the change of the phase of the standing wave potential. 
Other schemes of manipulating an optical superlattice for realizing the topological pumping have also been discussed.\cite{ChiangNiu, WangTroyerDai, Marra, WeiMueller} 

The integer QH effect and its 1D counterpart, the Thouless pumping, are composed of non-interacting free fermions. 
It is well known that in high magnetic fields, interactions between particles can generate highly entangled ground states and result in 
various QH states with fascinating features. 
For example, the fractional QH (FQH) state\cite{Tsui, Laughlin} and its non-Abelian generalizations\cite{MooreRead, ReadRezayi, NASS} exhibit fractionally quantized Hall conductivity and host fractionalized excitations with exotic quantum statistics which differs from ordinary bosons and fermions. The FQH states are examples of 
topologically ordered states,\cite{WenIntJMod} and exhibit ground-state degeneracy if the system is put on a topologically nontrivial surface with nonzero genus, such as a torus.\cite{NiuWen}

Other than the FQH states, strong correlations can create yet another interesting \textit{integer} QH state. 
The bosonic integer QH (BIQH) state,\cite{SenthilLevin} which is formed by two-component bosons with the total filling factor $\nu=1+1$, is such an example. Since noninteracting bosons form a Bose-Einstein condensate in the ground state, QH states of bosons inevitably require interactions. Although the BIQH state does not exhibit any fractionalized excitation or topological ground-state degeneracy, it is strictly distinguished from a trivial phase as long as the U$(1)$ symmetry associated with the conservation of the total particle number is preserved.\cite{LuVishwanath} 
In this sense, the BIQH state is an example of a 2D symmetry-protected topological (SPT) phase\cite{GuWen, Chen, SPTbook} of bosons, and there have been studies on its physical properties,\cite{Grover,LiuWen,YeWen,Geraedts,LiuMei} physical models realizing it (particularly in cold-atom setups),\cite{FurukawaUeda, WuJain, RegnaultSenthil, Grass, Sterdyniak, He, Fuji, ZengBIQH,HeGrusdt,Moller1,Moller2} exotic phase transitions out of it,\cite{Grover,LuLee,Fuji} its relationship to quantum spin liquid,\cite{Barkeshli,He2, He3} and its generalizations.\cite{LiuWen,YeWen,Lapa} 
The Hall response of the BIQH state is described by the effective Chern-Simons theory\cite{SenthilLevin,LuVishwanath} 
\begin{equation}
\mathcal{L}=-\frac{1}{4\pi}\varepsilon^{\mu\nu\lambda}(A_{1\mu}\partial_\nu A_{2\lambda}+A_{2\mu}\partial_\nu A_{1\lambda}),
\label{CS}
\end{equation}
where $A_{1\mu}$ $(A_{2\mu})$ is the U$(1)$ gauge field which couples to the first (second) component of bosons. Here we assume that the number of particles is separately conserved in each component. 
From the effective theory \eqref{CS}, we can read off that a probe electric field for \textit{one} component induces the quantized Hall response in \textit{the other}. Such an ``off-diagonal'' response is a unique feature of the BIQH state, which we highlight in this paper. 

In this paper, we aim to construct nontrivial classes of topological pumping which correspond to QH states created by strong interactions. We first consider topological pumping which mimics the Hall response of the FQH states, and then focus on the case of the BIQH state. Based on a quasi-one-dimensional (quasi-1D) limit of QH states,\cite{TaoThouless, Seidel, Bergholtz1, Bergholtz2} we can systematically construct strongly interacting models of topological pumping and thus naturally extend the connection between the topological pumping and the QH effect to interacting systems. We show that the quasi-1D limit of the BIQH state is given by the Haldane phase,\cite{HaldanePhysLett, Haldane, AKLT1, AKLT2} which is a celebrated example of a SPT phase in one dimension.\cite{GuWen, Pollmann1, Pollmann2} The mechanism of the resulting topological pumping is interpreted in terms of changes in polarizations between quantized values that correspond to two distinct gapped phases in the presence of the inversion symmetry. The obtained 
off-diagonal topological pumping intertwining two-component bosons provides novel interaction-induced topological pumping, and suggests an intriguing connection between the 2D topological phases and 1D gapped phases. 

The rest of this paper is organized as follows. In Sec.\ \ref{fluxins}, we describe our idea for obtaining models of topological pumping 
systematically from QH states on a thin torus. We take the case of FQH states as examples, and explain that their thin-torus counterparts naturally give rise to fractional charge pumping. In Sec.\ \ref{ttl}, we construct a 1D lattice model that corresponds to the thin-torus limit of two-component bosons in a magnetic field. We then identify the Haldane state as the thin-torus counterpart of the BIQH state. In Sec.\ \ref{Berryphase}, the topological pumping which mimics the BIQH effect is described and interpreted in terms of changes in polarizations between symmetry-protected quantized values. Finally, we summarize our results in Sec.\ \ref{summary}.

\section{Topological pumping as flux insertion through a thin torus\label{fluxins}}

\begin{figure}
\includegraphics[width=8.5cm]{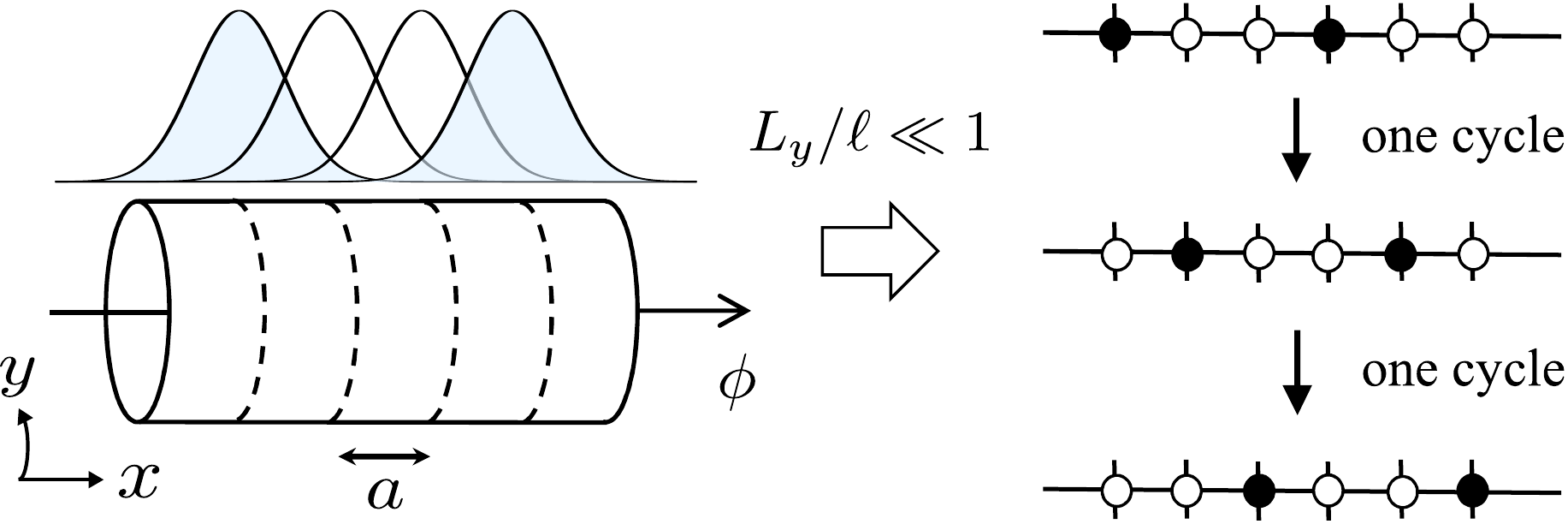}
\caption{
(color online) QH system on a thin torus (left) and topological pumping (right). A flux $\phi$ (in units of $\hbar/Q$) is inserted through a thin torus on which the LLL orbitals \eqref{LLLWF} are formed with a constant spacing $a$. In a thin-torus limit $L_y/\ell\ll 1$, the flux insertion argument for the Laughlin FQH states can be restated as the change between degenerate CDW ground states by a translation, which results in fractional Thouless pumping. 
The figure is the case of $\nu=1/3$.}
\label{fig_ttl}
\end{figure}

In this section, we explain our idea for obtaining models of 1D topological pumping systematically from a so-called thin-torus limit of 2D QH states.\cite{TaoThouless, Seidel, Bergholtz1, Bergholtz2} Let us consider a bosonic or fermionic system composed of $N$ particles of charge $Q$ and mass $M$ in a uniform magnetic field $B$ on a 2D torus of size $L_x\times L_y$. 
We take the Landau gauge $\bm{A}=(0,Bx)$ and assume $QB>0$. 
The total number of flux quanta piercing the system is $N_\phi=\frac{L_xL_y}{2\pi\ell^2}$, where $\ell=\sqrt{\frac{\hbar}{QB}}$ is the magnetic length. 
The filling factor is defined as $\nu=N/N_\phi$. 
The single-particle spectrum is given by the Landau levels $E_n=\hbar\Omega(n+\frac{1}{2})~(n=0,1,\dots)$, where $\hbar\Omega=\frac{\hbar^2}{2M\ell^2}$ is the cyclotron energy. 
The eigenstates in each level are $N_\phi$-fold degenerate, and labeled in the present gauge by the wave number $k_m=\frac{2\pi m}{L_y}$ $(m=0,1, \cdots, N_\phi-1)$ in the $y$ direction. 
In particular, the lowest-Landau-level (LLL) wave functions are given by\cite{Yoshioka,Haldane2} 
\begin{align}
\psi_m(\bmr)=\frac{1}{\sqrt{\pi^{1/2}\ell L_y}}\sum_{n\in\mathbb{Z}}\exp\Bigl[&-\frac{1}{2\ell^2}(x-k_m\ell^2-nL_x)^2\notag\\
&+i\Bigl(k_m+\frac{nL_x}{\ell^2}\Bigr)y\Bigr].
\label{LLLWF}
\end{align}
This wavefunction is localized around $x=k_m\ell^2= ma$ with a width $\ell$ in the $x$ direction, and delocalized in the $y$ direction (see the left panel of Fig.\ \ref{fig_ttl}).  
Here, $a=2\pi\ell^2/L_y=L_x/N_\phi$ is the spacing between neighboring wavefunctions, and used as an effective ``lattice constant'' later. 

In Laughlin's flux insertion argument,\cite{Laughlin2} the quantized Hall conductivity can be derived as a response to an adiabatic insertion of a flux quantum through the torus. 
The effect of flux insertion is expressed by a twisted boundary condition, which results in the replacement $k_m\to k_m+\frac{\phi}{L_y}$ with $0\leq\phi\leq2\pi$. After inserting one flux quantum ($\phi=2\pi$), the Hamiltonian of the system goes back to its original form. However, each Landau-level orbital shifts its position by $a$ in the $x$ direction during this process. 
Hence, if some of the Landau levels are completely occupied and others are empty, the system exhibits the integer QH effect. The FQH effect is also understood in a similar manner by taking into account the topological ground-state degeneracy on the torus.\cite{Haldane2,TaoWu, NiuWen} In this case, the insertion of one flux quantum transfers the initial ground state to another degenerate ground state, and some integer multiple of flux quanta are required to go back to the initial ground state.

Keeping this picture in mind, let us gradually decrease the length $L_y$ in the $y$ direction while keeping the total area $L_xL_y$ fixed. 
By identifying the locations $x=ma$ of the LLL orbitals \eqref{LLLWF} as a lattice coordinate, the original 2D system can be viewed as an effective 1D lattice model.\cite{TaoThouless, Seidel, Bergholtz1, Bergholtz2} The effective model contains some long-range interactions, whose coefficients are given by interaction matrix elements with respect to the LLL orbitals on concerned sites. 
However, as we decrease $L_y$, the lattice constant $a$ increases, and the LLL orbitals are mutually separated further. 
Interactions for longer distances are thus suppressed more rapidly in this process, and the physics for $L_y/\ell\ll 1$ is expected to be dominated by a few interaction terms for short distances. 

If the ground state of a given QH state is smoothly changed without closing an excitation gap with decreasing $L_y$, the ground state is expected to gradually acquire a 1D character owing to the suppression of interactions for long distances. We can then view the ground state on a thin torus as the 1D lattice counterpart of the QH state. It is then expected that the insertion of a flux quantum through the thin torus induces quantized current in the $x$ direction. This phenomenon can be viewed as the topological pumping in the 1D model.\cite{NiuThouless, NiuThoulessWu}

Let us illustrate the above idea using the simplest FQH state, the Laughlin state of fermions at $\nu=1/3$. It was shown that with decreasing $L_y$, the $\nu=1/3$ Laughlin state is smoothly deformed into a charge-density-wave (CDW) state in which every third site is occupied by a particle.\cite{Seidel} Here, density-density interactions for nearest-neighbor and next-nearest-neighbor 
pairs of sites play a dominant role and stabilize the CDW state. If a flux quantum is adiabatically inserted through a thin torus, 
the LLL orbitals shift their positions by one lattice spacing $a$, and the CDW ground state changes into another degenerate ground state\cite{NiuThoulessWu} as shown in Fig.\ \ref{fig_ttl}. 
The total current during this shift of the ground state is equal to $1/3$ if averaged in space; this corresponds to
the fractional Hall conductivity $\sigma_{xy}=\frac{1}{3}\frac{Q^2}{h}$ of the Laughlin state. Hence, as expected, the flux insertion for the 1D counterpart of the FQH state results in fractional Thouless pumping. Similarly to Laughlin's argument for the FQH effect,\cite{TaoWu} here the degeneracy of the ground states is essential in obtaining the fractional pumping. 

For a filling fraction with a denominator larger than $3$, it is not clear whether the FQH state is adiabatically connected to a CDW state with decreasing $L_y$. 
This is because in a thin torus, the density-density interactions that stabilize a CDW state can severely compete with other interaction terms in which two particles hop in a center-of-mass-conserving manner.\cite{Seidel,Bergholtz1,Bergholtz2,Nakamura, Wang} 
However, if we keep only the density-density interactions and neglect other interaction terms, the system exhibits a CDW ground state with $q$-fold degeneracy at every rational filing fraction $\nu=p/q$ (with $p$ and $q$ being coprime).\cite{Bergholtz1,Bergholtz2, Seidel, Rotondo} The adiabatic shift of such a CDW state by one lattice spacing clearly results in fractional Thouless pumping. 
Similar schemes for realizing fractional pumping by CDW states have been discussed in literature,\cite{Meidan, Grusdt, Zeng, Zeng2, Taddia} especially in connection with the ``synthetic dimension'' technique\cite{Celi} in cold atoms. 
In this technique, infinite-range interactions in the synthetic dimension stabilize CDW ground states.\cite{Grusdt, Zeng, Taddia, Barbarino, SaitoFurukawa} Although the topological pumping constructed here is just a translational operation of the entire system and seems to be somewhat trivial, we will see that the case of the BIQH state provides more nontrivial topological pumping.

\section{Thin-torus limit of the bosonic integer quantum Hall state\label{ttl}}

Based on the correspondence between the QH state and the topological pumping described in the previous section, we here construct the thin-torus counterpart of the BIQH state. Let us start with two-component bosons in a uniform magnetic field on a torus described by the Hamiltonian 
\begin{equation}\label{Hamil}
\begin{split}
H&=\sum_{\alpha=1,2}\int d^2\bmr\Psi^{(\alpha)\dag}(\bmr)\frac{[\bm{p}-Q\bm{A}(\bmr)]^2}{2M}\Psi^{(\alpha)}(\bmr)\\
&+\sum_{\alpha,\beta}\frac{g^{(\alpha\beta)}}{2}\int d^2\bmr \Psi^{(\alpha)\dag}(\bmr)\Psi^{(\beta)\dag}(\bmr)\Psi^{(\beta)}(\bmr)\Psi^{(\alpha)}(\bmr),
\end{split}
\end{equation}
where $\Psi^{(\alpha)}(\bmr)\ (\alpha=1,2)$ denotes the bosonic field operator for the $\alpha$-th component. 
We assume repulsive contact interactions $g^{(\alpha\beta)}>0$ between particles, and set $g^{(11)}=g^{(22)}\equiv g$ for simplicity. 
The filling factor for each component is set to unity so that the total filling factor is given by $\nu=1+1$. 
The system possesses the U(1)$\times$U(1) symmetry associated with the particle number conservation in each component. 
Through exact diagonalization analyses,\cite{FurukawaUeda, WuJain, RegnaultSenthil} it was shown that the BIQH state described by the effective Chern-Simons theory \eqref{CS} appears when the ratio of the intercomponent to intracomponent interactions, $\delta\equiv g^{(12)}/g$, is close to unity. 

Within the LLL approximation, the field operators are expanded as
\begin{equation}
\Psi^{(\alpha)}(\bmr)=\sum_{m=0}^{N_\phi-1}b^{(\alpha)}_m\psi_m(\bmr,\phi_\alpha),
\label{LLLexpansion}
\end{equation}
where $b_m^{(\alpha)}$ annihilates a particle in the $m$-th LLL orbital and satisfies the commutation relations $[b_m^{(\alpha)}, b_n^{(\beta)\dag}]=\delta_{\alpha\beta}\delta_{mn}$ and $[b_m^{(\alpha)}, b_n^{(\beta)}]=0$. 
To facilitate later discussions on topological pumping, we have introduced a magnetic flux $\phi_\alpha$ which couples to the $\alpha$-th component so that the LLL wave function $\psi_m(\bmr, \phi_\alpha)$ is given by the right-hand side of Eq.\ (\ref{LLLWF}) 
with $k_m$ replaced by $k_m+\frac{\phi_\alpha}{Ly}$. Correspondingly, the positions of the LLL orbitals are shifted to $x=(m+\frac{\phi_\alpha}{2\pi})a$ in the $\alpha$-th component. 
Substituting the expansion (\ref{LLLexpansion}) into the Hamiltonian (\ref{Hamil}), we obtain
\begin{align}
&H=\frac{1}{2}\hbar\Omega \sum_\alpha N^{(\alpha)}\notag\\
&+\sum_{\alpha}\sum_j\sum_{|n|\leq m\leq \frac{N_\phi}{2}}V_{mn}b_{j+n}^{(\alpha)\dag}b_{j+m}^{(\alpha)\dag}b_{j+m+n}^{(\alpha)}b_{j}^{(\alpha)}\notag\\
&+\sum_j\sum_{-\frac{N_\phi}{2}<m,n\leq \frac{N_\phi}{2}}V^{(12)}_{mn}(\phi_1-\phi_2)b_{j+n}^{(1)\dag}b_{j+m}^{(2)\dag}b_{j+m+n}^{(2)}b_{j}^{(1)}.\label{H_LLL}
\end{align}
We note that the interaction in this Hamiltonian preserves the center-of-mass position in the $x$ direction. 
The interaction matrix elements are calculated by using the LLL wave functions as \cite{Yoshioka,Bergholtz2}
\begin{subequations}
\begin{align}
&V_{mn}=\frac{z_{mn} g}{L_xL_y}\sum_{\bmq}\left[ \delta_{n,n_y}' e^{-\frac{1}{2}\bmq^2\ell^2}\cos(q_xk_m\ell^2)+(m\leftrightarrow n) \right],\label{V3}\\
&V^{(12)}_{mn}(\phi)=\frac{g^{(12)}}{L_xL_y}\sum_{\bmq} \delta_{n,n_y}' e^{-\frac{1}{2}\bmq^2\ell^2}\cos\left[q_x \left(k_m-\frac{\phi}{L_y}\right)\ell^2\right],\label{V3_12}
\end{align}
\end{subequations}
where the sum is over the wave vector $\bm{q}=(\frac{2\pi n_x}{L_x}, \frac{2\pi n_y}{L_y})$ $(n_x, n_y\in\mathbb{Z})$, $\delta'_{n,n_y}$ is the modulo-$N_\phi$ Kronecker delta, and $\displaystyle z_{mn}=2^{-\delta_{m,|n|}(1+\delta_{m,0})}2^{-\delta_{m,N_\phi/2}(1+\delta_{|n|,N_\phi/2})}$ is a factor for fixing the double counting of some terms. 
If we take the limit $L_x/\ell\to\infty$ (and thus $N_\phi\to\infty$) while keeping $L_y/\ell$ fixed, these elements are given more simply by 
\begin{subequations}\label{Vmn_simple}
\begin{align}
&V_{mn}=\frac{2z_{mn}g}{\sqrt{2\pi} L_y\ell} e^{-\frac12 (k_m^2+k_n^2)\ell^2}, \\
&V_{mn}^{(12)}(\phi)=\frac{g^{(12)}}{\sqrt{2\pi} L_y\ell} e^{-\frac12 [(k_m-\phi/L_y)^2+k_n^2]\ell^2}.
\end{align}
\end{subequations}

Let us first consider the case when no flux is inserted through the torus: $\phi_1=\phi_2=0$. When we take the thin-torus limit $L_y\to 0$, the only remaining interactions are on-site ones, which are the $m=n=0$ components of Eq.\ \eqref{H_LLL}. We thus obtain
\begin{align}
H=&\sum_{\alpha=1,2}\sum_j V_{00}n^{(\alpha)}_j(n^{(\alpha)}_j-1)
+\sum_j V^{(12)}_{00}n^{(1)}_j n^{(2)}_{j},\label{H_ttl}
\end{align}
where we ignore the constant kinetic energy of the LLL. 
The ground state of the thin-torus Hamiltonian (\ref{H_ttl}) is easily obtained. For $\delta<1$, where the intracomponent interaction is dominant ($2V_{00}>V^{(12)}_{00}$), the ground state is the product state of Bose Mott insulators
\begin{equation}
\Bigl|\begin{matrix}
\cdots\ n_j^{(1)}\ \cdots\\
\cdots\ n_j^{(2)}\ \cdots
\end{matrix}
\Bigr\rangle=
\Bigl|\begin{matrix}
\cdots\ 1\ 1\ 1\ 1\ \cdots\\
\cdots\ 1\ 1\ 1\ 1\ \cdots
\end{matrix}\Bigr\rangle. 
\label{Mott}
\end{equation}
For $\delta>1$, where the intercomponent interaction is dominant ($2V_{00}<V^{(12)}_{00}$), the ground states are ferromagnetic states
\begin{equation}
\Bigl|\begin{matrix}
\cdots\ n_j^{(1)}\ \cdots\\
\cdots\ n_j^{(2)}\ \cdots
\end{matrix}
\Bigr\rangle=
\Bigl|\begin{matrix}
\cdots\ 2\ 2\ 2\ 2\ \cdots\\
\cdots\ 0\ 0\ 0\ 0\ \cdots
\end{matrix}\Bigr\rangle
,~~
\Bigl|\begin{matrix}
\cdots\ 0\ 0\ 0\ 0\ \cdots\\
\cdots\ 2\ 2\ 2\ 2\ \cdots
\end{matrix}\Bigr\rangle,
\label{FM}
\end{equation}
if we fix only the total number of particles. 
If we fix the number of particles in each component, a phase separation occurs. 
The point $\delta=1$, at which $2V_{00}=V^{(12)}_{00}$, is special---the on-site energy is the same for $\ket{n^{(1)}_j,n^{(2)}_j}=\ket{2,0}, \ket{1,1}, \ket{0,2}$, leading to $3^{N_\phi}$-fold degeneracy of the ground state. 
This macroscopic degeneracy in the thin-torus limit is lifted by fluctuations as we increase $L_y$.

To obtain a unique ground state at $\delta=1$, we proceed away from the thin-torus limit by increasing $L_y$ and consider leading fluctuations. The next-leading interactions are the  nearest-neighbor ones, which are $V_{10}, V^{(12)}_{10} (=V^{(12)}_{-1,0})$, and $V^{(12)}_{01}$ terms. 
The $V^{(12)}_{01}$ term involves hopping of particles, while the other ones are of an electrostatic type. 
To discuss the competition of these terms, we restrict ourselves to the low-energy manifold of the Hilbert space spanned by the $3^{N_\phi}$-fold degenerate ground states of Eq.~\eqref{H_ttl} at $\delta=1$. 
The thin-torus ground states \eqref{Mott} and \eqref{FM} for $\delta\ne 1$ also reside in this manifold. 
In this restricted subspace, in which the constraint $\sum_\alpha n_j^{(\alpha)}=2$ is satisfied at every site, the operators
\begin{equation}
 \bm{S}_j = \frac12 \sum_{\alpha,\beta} b_j^{(\alpha)\dagger} \bm{\sigma}_{\alpha\beta} b_j^{(\beta)} 
\end{equation}
satisfy the commutation relations of the SU(2) generators and have the fixed magnitude $\bm{S}_j^2=1(1+1)$ as is known in the Schwinger boson formalism.\cite{Auerbach} 
Here, $\bm{\sigma}=(\sigma^x,\sigma^y,\sigma^z)$ is a set of Pauli matrices. 
The Hamiltonian can thus be written in terms of the spin-$1$ operators as 
\begin{align}
H=\sum_j \left[J_{xy}(S^x_jS^x_{j+1}+S^y_jS^y_{j+1})+J_zS^z_jS^z_{j+1}+D(S^z_j)^2 \right],\label{H_XXZ}
\end{align}
where $J_{xy}=2V^{(12)}_{01}$, $J_z=2(V_{10}-V^{(12)}_{10})$, and $D=2V_{00}-V^{(12)}_{00}$. This has the form of the XXZ chain with a single-ion anisotropy.\cite{denNijs,KennedyTasaki,ChenHidaSanctuary} At $\delta=1$, in particular, since $V_{10}=2V^{(12)}_{10}=2V^{(12)}_{01},$ the effective Hamiltonian is given by the spin-$1$ antiferromagnetic Heisenberg chain
\begin{align}
H=J\sum_j\bm{S}_j\cdot\bm{S}_{j+1},
\label{Heis}
\end{align}
where we set $V_{10}\equiv J>0$. 
This Hamiltonian has a non-degenerate ground state known as the Haldane state.\cite{HaldanePhysLett, Haldane, AKLT1, AKLT2} The macroscopic degeneracy of the ground state of Eq.\ \eqref{H_ttl} is thus lifted by the leading fluctuations.

\begin{figure}
\includegraphics[width=7.5cm]{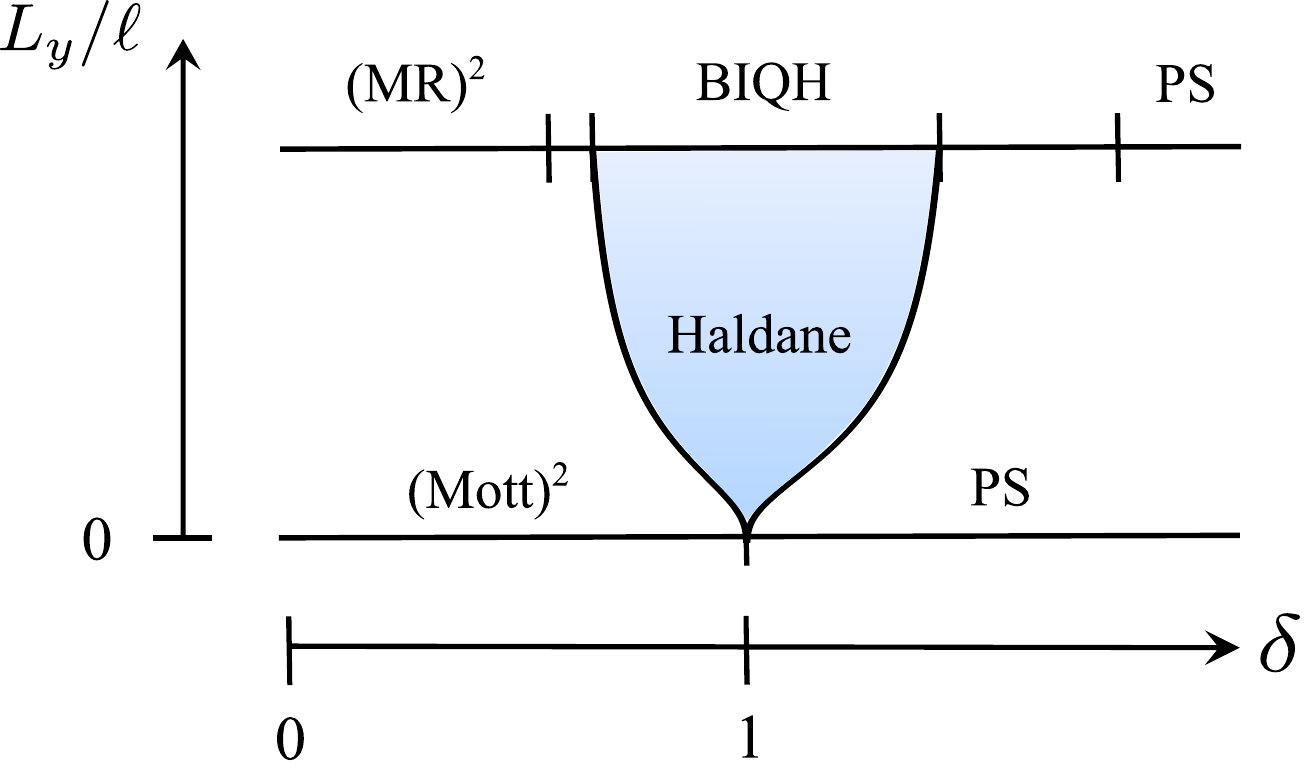}
\caption{(color online) Schematic phase diagram of two-component bosons (\ref{Hamil}) at $\nu=1+1$ in the space of the ratio of the intercomponent to intracomponent interactions, $\delta=g^{(12)}/g$, and the length $L_y$ in the $y$ direction. 
A quasi-1D limit $L_y/\ell\ll 1$ is described by the spin-$1$ chain \eqref{H_XXZ} 
while the 2D phase diagram has been studied in Refs.~\onlinecite{RegnaultSenthil, FurukawaUeda2}. 
The product of Bose Mott insulators [(Mott)$^2$] in the quasi-1D limit is expected to evolve into the the product of Moore-Read states [(MR)$^2$] for small $\delta$ in the 2D case. 
A phase separation (PS) occurs for $\delta>1$ in the quasi-1D limit and for $\delta\gtrsim 2.5$ in the 2D case. 
A Haldane phase that intervenes between the regions of (Mott)$^2$ and PS in the quasi-1D limit is expected to evolve into the BIQH phase in the 2D system. }
\label{fig_phase}
\end{figure}

At this stage, it is interesting to compare the phases of the spin-$1$ chain (\ref{H_XXZ})\cite{denNijs,KennedyTasaki,ChenHidaSanctuary} with the phase diagram of the original 2D system. 
The two-component bosons (\ref{Hamil}) at $\nu=1+1$ in two dimensions show a couple of phases when varying the interaction ratio $\delta$ (Fig.\ \ref{fig_phase}).\cite{RegnaultSenthil, FurukawaUeda2} When $\delta$ is small, the two components are nearly decoupled and form the Moore-Read states\cite{Cooper} independently. In the opposite limit, when $\delta$ is large (numerically $\delta\gtrsim2.5$), the 2D ground state exhibits a phase separation.  
The BIQH state appears around $\delta=1$ intervening the two limiting cases.\cite{FurukawaUeda, WuJain, RegnaultSenthil}  
Our mapping to a spin chain (\ref{H_XXZ}) qualitatively reproduces these phases as summarized in Fig.\ \ref{fig_phase}. 
When $L_y/\ell$ is sufficiently small, the single-ion anisotropy $D$ is the most dominant term in Eq.\ (\ref{H_XXZ}). 
For $\delta< 1$, we have $D>0$, and the ground state is a large-$D$ state, which is equivalent to the doubled Mott insulators in Eq.\ \eqref{Mott}. 
For a single-component Bose gas at the filling factor $\nu=1$, it was shown that the ground state in the thin-torus limit is given by a Bose Mott insulator $\ket{\cdots\ n_j^{(\alpha)}\ \cdots}=\ket{\cdots\ 1\ 1\ 1\ 1\ \cdots}$, and then two CDW states 
$\ket{\cdots\ n_j^{(\alpha)}\ \cdots}=\ket{\cdots\ 2\ 0\ 2\ 0\ \cdots}$ and $\ket{\cdots\ 0\ 2\ 0\ 2\ \cdots}$ become nearly degenerate with the ground state as we increase $L_y/\ell$; these three states naturally evolve into the threefold degenerate ground states of the bosonic Moore-Read state on a 2D torus.\cite{Wikberg} Similarly, the doubled Mott insulators in Eq.\ \eqref{Mott} are expected to evolve into the doubled Moore-Read states found in the 2D system. 
For $\delta>1$, we have $D<0$ and $J_z<0$, and the spin chain \eqref{H_XXZ} exhibits ferromagnetic ground states as in Eq.\ \eqref{FM}; if the total magnetization of the system is fixed at zero, a phase separation occurs as found in the 2D system. 
As we increase $L_y/\ell$, the Haldane phase appears between the large-$D$ and ferromagnetic phases in the spin-chain model, and its range along the $\delta$ axis gradually increases. 
Owing to the uniqueness of the ground state and high entanglement between the two components, it is natural to speculate that this phase evolves into the BIQH phase in the 2D case. 
It is worth noting that our mapping to a spin chain can be generalized to the case of arbitrary integer $\nu$, resulting in a spin-$\frac{\nu}{2}$ version of Eq.~\eqref{H_XXZ}. 
Then the Haldane conjecture for the Heisenberg chain \eqref{Heis} \cite{HaldanePhysLett, Haldane} suggests the emergence of gapped (gapless) states for even (odd) $\nu$. 
The gapless state at $\nu=1/2+1/2$ is expected to evolve into a gapless composite fermion liquid in the 2D case.\cite{WuJain2}

Next, let us consider the thin-torus limit in the case when some fluxes $\phi_\alpha$ are inserted through the torus. 
Since the Hamiltonian (\ref{H_LLL}) depends on the fluxes only through $\phi_1-\phi_2$, we set $\phi_1\neq 2\pi n$ ($n\in\mathbb{Z}$) and $\phi_2=0$ without loss of generality. In this case, since the flux causes lattice mismatch between the two components, 
the only remaining interaction in the limit $L_y/\ell \to 0$ is the on-site intracomponent one $V_{00}$, and the intercomponent interactions completely disappear. 
The ground state is thus the doubled Mott insulators (\ref{Mott}) at any $\delta$. We note that this fact does not contradict the above identification of the BIQH state with the Haldane phase, since the flux insertion breaks the inversion symmetry of the system (except at $\phi_1=\pi$) and thus there is no clear distinction between the Haldane phase and the Bose Mott insulator.\cite{DallaTorre, Berg, BergLevinAltman} 
In the next section, we show that the change in the ground state during the adiabatic flux insertion is related to the Hall response of the BIQH state, leading to the off-diagonal topological pumping in the quasi-1D system.

\section{Off-diagonal topological pumping in the thin-torus limit \label{Berryphase}}

In this section, we describe the 1D topological pumping which mimics the BIQH effect, and thereby reinterpret the fact that the thin-torus counterpart of the BIQH state is the Haldane phase. 
We start from the two-component Bose system \eqref{Hamil} on a 2D torus, and consider its Hall response. 
As discussed in Sec.\ \ref{ttl}, we introduce a magnetic flux $\phi_\alpha~(\alpha=1,2)$ through the torus, which results in the twisted boundary condition for the $\alpha$-th component in the $y$ direction. 
We also introduce a magnetic flux $\theta_\alpha~(\alpha=1,2)$ through the other direction of the torus, which results in the analogous twisted boundary condition in the $x$ direction. 
Then the Hall response of the $\alpha$-th component to the flux insertion for the $\beta$-th component ($\alpha, \beta=1,2$) 
can be expressed by the many-body Chern number\cite{NiuThoulessWu} 
\begin{equation}
C_{\alpha\beta}=\frac{1}{2\pi i}\int_0^{2\pi}\!\!\! d\theta_\alpha\int_0^{2\pi}\!\!\! d\phi_\beta (\braket{\partial_{\theta_\alpha}\psi|\partial_{\phi_\beta}\psi}-\braket{\partial_{\phi_\beta}\psi|\partial_{\theta_\alpha}\psi}),
\label{Chern}
\end{equation}
where $\ket{\psi(\theta_\alpha,\phi_\beta)}$ is the many-body ground state for the twists $\theta_\alpha$ and $\phi_\beta$ for the $\alpha$-th and $\beta$-th components in the $x$ and $y$ directions, respectively, while the other twisting angles are set to zero. 
The Chern-Simons theory (\ref{CS}) of the BIQH effect corresponds to $C_{11}=C_{22}=0$ and $C_{12}=C_{21}=1$. 
These responses result in the off-diagonal topological pumping in the thin-torus limit as we see below. 

Hereafter we focus on the responses corresponding to $C_{11}$ and $C_{21}$. 
These can be analyzed by setting $\phi_2=0$ and adiabatically changing the pumping parameter $t\equiv\frac{\phi_1}{2\pi}T$ from $0$ to $T$. The Hall current in the $x$ direction is identified as the pumped charge. 
For fixed $t$, it is useful to introduce the Berry phase
\begin{equation}
\gamma_\alpha(t)=-\int_0^{2\pi}d\theta_{\alpha}\bra{\psi(\theta_\alpha,t)}\partial_{\theta_\alpha}\ket{\psi(\theta_\alpha,t)} \ \  (\mathrm{mod}\ 2\pi),
\label{Berry}
\end{equation}
which is associated with the change in the ground state $\ket{\psi(\theta_\alpha, t)}$ when $\theta_\alpha$ is adiabatically changed from $0$ to $2\pi$. 
The Chern number (\ref{Chern}) can then be rewritten as
\begin{equation}
C_{\alpha,\beta=1}=-\frac{1}{2\pi}\int_0^{T}dt~ \partial_{t}\gamma_\alpha(t).
\label{Chern2}
\end{equation}
In this expression, the quantized Hall response, or equivalently the quantized charge pumping can be understood as 
$2\pi n$ ($n\in \mathbb{Z}$) change in the Berry phase $\gamma_\alpha(t)$ over the pumping process.\cite{Zaletel}

A more intuitive understanding of the quantized pumping can be gained by introducing the polarization. 
To introduce it, we define 
\begin{align}
 z_\alpha(t) = \bra{\psi(t)}\exp \left[ \frac{2\pi i}{L_x} \int d\bm{r}~x \Psi^{(\alpha)\dagger}(\bm{r}) \Psi^{(\alpha)}(\bm{r}) \right] \ket{\psi(t)}, 
\end{align}
where $\ket{\psi(t)}=\ket{\psi(\theta_\alpha=0,t)}$. 
This is convenient in describing the center-of-mass position of the particles in the $\alpha$-th component in the $x$ direction
since the position $x$ is defined modulo $L_x$ under the periodic boundary condition. 
Within the LLL approximation, we can exploit the fact that the $j$-th LLL orbital for the $\alpha$-th component is localized around $x_\alpha(j)=(j+\frac{t}{T}\delta_{\alpha, 1})a$ in the $x$ direction (with larger spacing $a$ for smaller $L_y/\ell$), and approximate $z_\alpha(t)$ as
\begin{align}
z_\alpha(t)
\approx &\bra{\psi(t)}\exp\Bigl[\frac{2\pi i}{N_\phi a}\sum_j x_\alpha(j)n^{(\alpha)}_j\Bigr]\ket{\psi(t)}\notag\\
=&\bra{\psi(t)}\exp\Bigl[\frac{2\pi i}{N_\phi}\sum_j jn^{(\alpha)}_j+\frac{2\pi it}{T}\delta_{\alpha,1}\Bigr]\ket{\psi(t)}. \label{z_LLL} 
\end{align}
This resembles the expectation value of the Lieb-Schultz-Mattis twist operator.\cite{LSM, AffleckLieb, OYA, NakamuraTodo, NakamuraVoit, Oshikawa} 
The phase of $z_\alpha(t)$ gives the polarization\cite{Resta2,RestaSorella}
\begin{equation}
 P_\alpha(t)=\frac{1}{2\pi}\mathrm{Im}\ln z_\alpha(t) \ \ (\mathrm{mod}\ 1).
\end{equation}
Importantly, the polarization is directly related to the Berry phase as\cite{Resta1, KingSmith, OrtizMartin, Resta2} 
\begin{equation}
P_\alpha(t)=-\frac{1}{2\pi}\gamma_\alpha(t).
\end{equation}
The Chern number can therefore be rewritten as the change in the polarization over the pumping cycle: 
\begin{equation}
C_{\alpha,\beta=1}=\Delta P_\alpha=\int_0^{T}dt~ \partial_{t}P_\alpha(t).
\label{Chern2}
\end{equation}
The BIQH effect should thus correspond to the topological pumping with $\Delta P_1=0$ and $\Delta P_2=1$. 

Let us now discuss in detail the topological pumping in the thin-torus limit of the BIQH state. 
The pumping protocol in the present case is not just the translation but involves the change of intercomponent interactions, 
in sharp contrast with the FQH cases discussed in Sec. \ref{fluxins}. 
Since the Hamiltonian \eqref{H_LLL} in the LLL basis is invariant under the combined operations of the spatial inversion $j\to N_\phi-j$ and the interchange of two components $1\leftrightarrow 2$, $z_\alpha(t)$ in Eq.\ \eqref{z_LLL} satisfies
\begin{equation}
 z_1(t) = e^{2\pi it/T} z_2^*(t),
\end{equation}
which indicates $P_1(t)+P_2(t)=t/T$. Thus, $\Delta P_1=0$ implies $\Delta P_2=1$ and vice versa. Furthermore, by exploiting 
the invariance of the pumping protocol under the combined operations of the spatial inversion and time reversal $t\to -t$, 
we find $P_\alpha(-t)=-P_\alpha (t)$, from which we obtain $\Delta P_\alpha=2\int_0^{T/2}dt~\partial_tP_\alpha(t)$. 
Thus, half of the expected changes $\Delta P_\alpha=\delta_{\alpha,2}$ in the polarizations over one cycle must occur during $t\in [0,T/2]$: 
\begin{equation}\label{P_half}
P_\alpha(T/2)-P_\alpha(0)=\frac12 \delta_{\alpha,2}. 
\end{equation}

To discuss the variation of the polarizations $P_\alpha (t)$ as functions of $t$, 
it is important to notice that the system possesses the spatial inversion symmetry at $t=0$ and $t=T/2$. Therefore, the polarization $P_\alpha$ must be quantized to $0$ or $1/2$ at these values of $t$. 
It is known that in the presence of the inversion symmetry, a change in the polarization between these quantized values in general signals a phase transition.\cite{NakamuraTodo,NakamuraVoit} 
The polarization can therefore be used as an order parameter for detecting 1D topological phases protected by the inversion symmetry.\cite{NakamuraTodo,NakamuraVoit,Qi, Hughes} 
Similar results can also be obtained through the quantization of the Berry phase.\cite{Zak, Hatsugai, Hirano,  Kariyado1, Kariyado2} 
At $t=0$, the thin-torus limit of the BIQH state is given by the Haldane state as discussed in Sec.\ \ref{ttl};  
the Haldane phase is known to be a topological phase protected by the inversion symmetry,\cite{GuWen, Pollmann1, Pollmann2} and has the polarizations\cite{NakamuraTodo,NakamuraVoit} 
\begin{equation}\label{P_Hal}
 P_1(0)=P_2(0)=N_\phi/2.
\end{equation}
At $t=T/2$, the ground state in the thin-torus limit is given by the doubled Mott insulators;  a direct calculation using Eq.\ \eqref{Mott} yields 
\begin{equation}\label{P_Mott}
 P_1(T/2)=N_\phi/2,~~P_2(T/2)=(N_\phi+1)/2.
\end{equation} 
The Haldane phase and the doubled Mott insulators have the different polarizations \eqref{P_Hal} and \eqref{P_Mott}, and thus are distinct phases as long as the system possesses the inversion symmetry. The topological pumping can be interpreted as a process of connecting between the two phases smoothly by breaking the inversion symmetry. 
The idea of utilizing symmetry-breaking perturbations to connect between otherwise distinct gapped phases has also been used in other examples of topological pumping such as those based on the Su-Schrieffer-Heeger model,\cite{RiceMele, NiuReview, SSH} the spin-Peierls phases\cite{Shindou} (equivalent to an interacting Su-Schrieffer-Heeger model via the Jordan-Wigner transformation), and the Haldane insulator phase of an extended Bose-Hubbard model.\cite{BergLevinAltman} 
We note that the polarizations in Eqs.\ \eqref{P_Hal} and \eqref{P_Mott} are consistent with the relations \eqref{P_half}; conversely, the relations \eqref{P_half} require the appearance of topologically distinct phases (in an inversion-symmetry-protected sense) at $t=0$ and $T/2$. 
This gives an explanation on why the Haldane phase, a typical example of a 1D SPT phase, should emerge in the thin-torus limit of the BIQH state. 

\begin{figure}
\includegraphics[width=8.5cm]{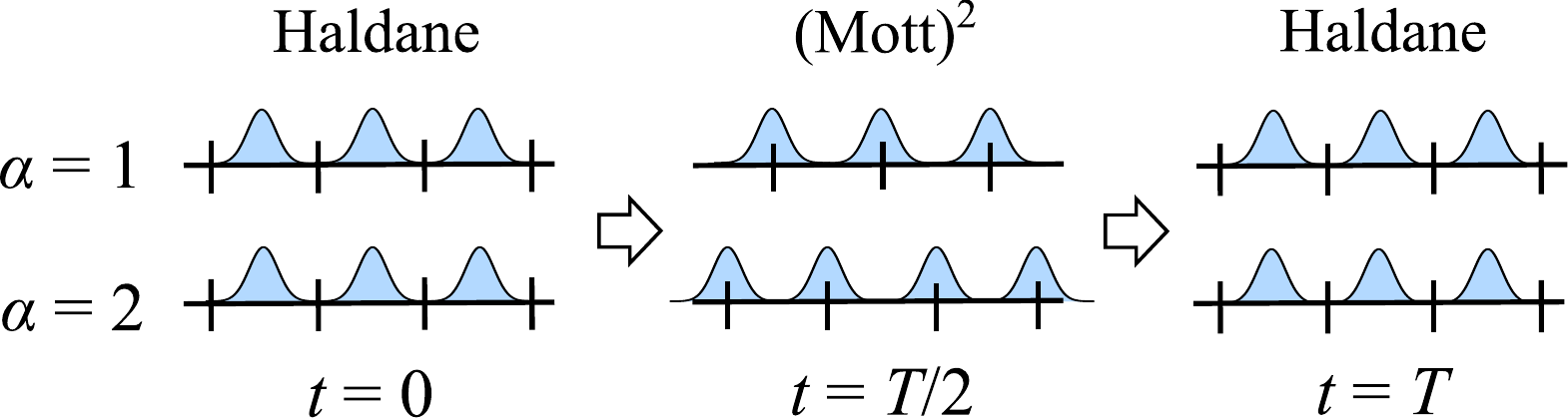}
\caption{(color online) Schematic picture of the off-diagonal topological pumping in the thin-torus limit of the BIQH state. 
The upper and lower chains describe the first and second components, respectively. }
\label{fig_pump}
\end{figure}

Figure \ref{fig_pump} summarizes the process of the off-diagonal topological pumping in the thin-torus limit. 
The Haldane phase at $t=0$ can intuitively be described as localization of bosons at bonds of the lattice in a way similar to that of one-component bosons.\cite{Berg} By shifting the lattice for the first component, the inversion symmetry is broken and the ground state is smoothly deformed into the doubled Mott insulators at $t=T/2$, and finally returns to the original ground state at $t=T$. During this process, the bosons in the second component are pumped by one lattice spacing while the bosons in the first component stay around the same positions.

Two remarks are in order. 
First, although the Haldane phase in the spin-$1$ chain can be protected not only by the inversion symmetry but also by the time-reversal or the spin rotation symmetry,\cite{GuWen, Pollmann1, Pollmann2} the latter two symmetries do not protect the Haldane phase in the case of soft-core bosons as in the present case.\cite{DallaTorre, Berg, BergLevinAltman} 
This is because the spin-$1$ degrees of freedom required for the symmetry protection argument are not perfectly formed in the presence of fluctuations in on-site particle numbers.  
The inversion symmetry is thus the only symmetry that protects the Haldane phase at $t=0$ in the present case. 
The second remark is on the mapping to a spin-$1$ chain done in Sec.~\ref{ttl}. In the mapping from the thin-torus Hamiltonian to a spin chain in the FQH cases, the inversion symmetry is sometimes broken since the mapping process involves grouping of several neighboring sites starting from CDW ground states.\cite{Nakamura, Wang} In the case of the BIQH state, in contrast, the mapping to a spin chain retains the inversion symmetry of the original system, since each spin-$1$ degree of freedom is composed of bosons at the same site. 

\begin{figure}
\includegraphics[width=8.cm]{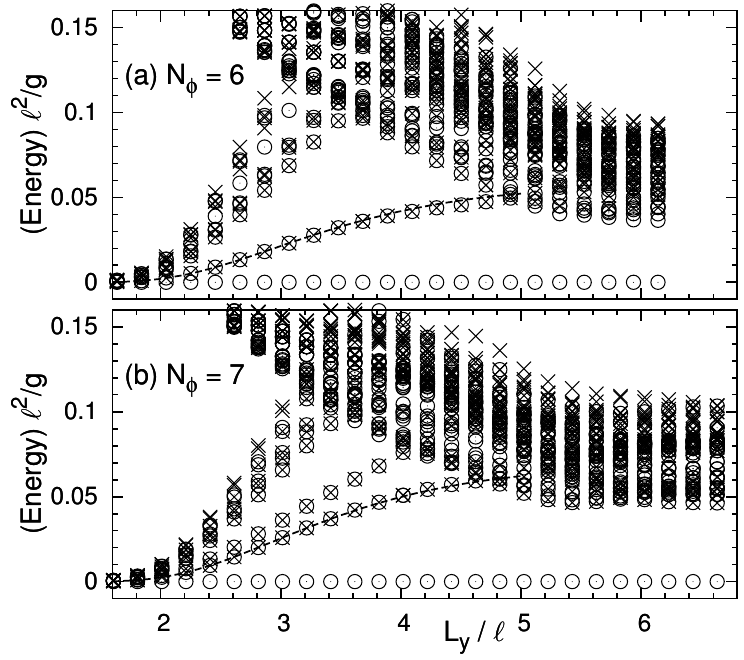}
\caption{Energy spectrum of the Hamiltonian \eqref{H_LLL} as a function of $L_y/\ell$ for (a) $N_\phi=6$ and (b) $N_\phi=7$. 
The ground-state energy is subtracted from the spectrum. 
Circles indicate eigenenergies in the equal-population case $N_1=N_2=N_\phi$. 
Crosses indicates eigenenergies in the minimally imbalanced case $(N_1,N_2)=(N_\phi+1,N_\phi-1)$. 
The two lowest energies for each pseudomomentum are shown. 
The data for the largest $L_y/\ell$ correspond to the case of $L_x=L_y$. 
Dashed lines indicate the energy gaps $0.721 J$ and $0.857 J$ of the spin-$1$ Heisenberg chain \eqref{Heis} with $N_\phi=6$ and $7$ spins, respectively, where $J$ is given by $V_{10}=2V_{10}^{(12)}$ in Eq.~\eqref{Vmn_simple} (we note that the gap in the thermodynamic limit\cite{WhiteHuse,Golinelli} is given by $0.410 J$). 
}
\label{fig_spec_Ly}
\end{figure}

\begin{figure}
\includegraphics[width=8.5cm]{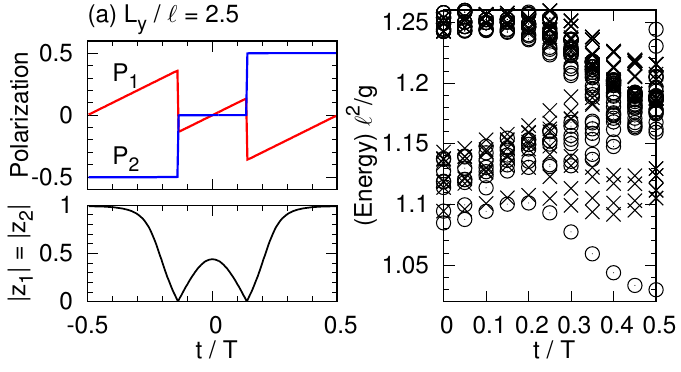}
\includegraphics[width=8.5cm]{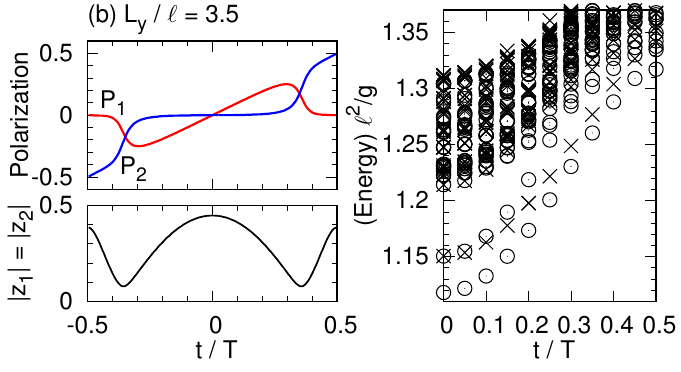}
\includegraphics[width=8.5cm]{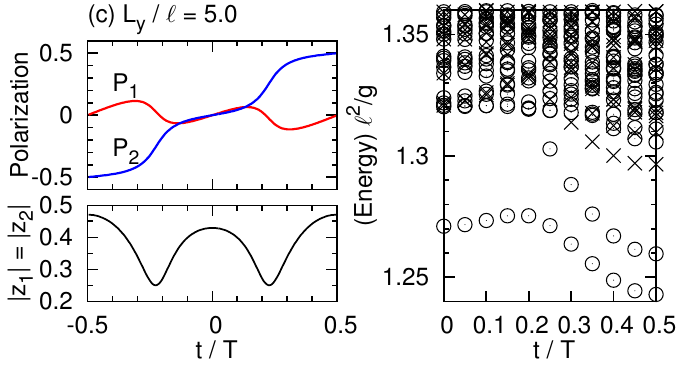}
\caption{(color online) The polarizations $P_\alpha$ (left top), the amplitudes $|z_1|=|z_2|$ of the twist operators (left bottom), and the energy spectrum (right) as functions of the pumping parameter $t\in [-T/2,T/2)$, for $N_\phi=6$ and (a) $L_y/\ell=2.5$, (b) 3.5, and (c) 5.0. 
Circles and crosses indicates eigenenergies in the equal-population and minimally imbalanced cases, respectively, as in Fig.~\ref{fig_spec_Ly}. 
Since the energy spectrum is symmetric around $t=0$, it is shown only for $t\in [0,T/2]$. 
}
\label{fig_pol_phi}
\end{figure}

To support the above picture of topological pumping, we have performed exact diagonalization calculations for the Hamiltonian \eqref{H_LLL} with the number of flux quanta up to $N_\phi=7$. 
We consider contact interactions with $g^{(12)}=g$. Figure \ref{fig_spec_Ly} presents the energy spectrum as a function of $L_y/\ell$. 
The ground state is found to remain in the sector with zero pseudomomentum, indicating that the BIQH state in two dimensions is smoothly deformed into the Haldane state in the thin-torus limit. 
For $L_y/\ell\lesssim 5$, the energy gap above the ground state agrees well with the finite-size energy gap of the spin-$1$ Heisenberg chain \eqref{Heis} (dashed lines) calculated by KOBEPACK.\cite{KOBEPACK,Kaburagi} 
Reflecting the large ground-state degeneracy in the thin-torus limit, a large number of eigenenergies collapse onto the ground-state energy with decreasing $L_y/\ell$. 
For $L_y/\ell\gtrsim 5$, the energy gap stays around a constant value, indicating the convergence to a 2D system. Figure \ref{fig_pol_phi} presents the polarizations $P_\alpha$, the amplitudes $|z_1|=|z_2|$ of the twist operators, and the energy spectrum as functions of the pumping parameter $t$. 
The ground state is found to remain in the zero-pseudomomentum sector in this process also. 
The calculated polarizations smoothly connect between Eqs.\ \eqref{P_Hal} and \eqref{P_Mott} for the Haldane state ($t=0$) and the doubled Mott insulators ($t=\pm T/2$). 
While the polarization of the first component stays around $P_1=0$, the second component shows $\Delta P_2=1$ over the cycle, clearly signaling the off-diagonal topological pumping. 
As we decrease $L_y/\ell$, long-range interactions are suppressed and the system gradually acquires a 1D character. 
Correspondingly, the change in the polarization becomes sharper in the quasi-1D limit. A rapid change in the center-of-mass position (and thus the polarization) has also been observed in 1D topological pumping.\cite{Nakajima, Lohse, WangTroyerDai, WeiMueller} 
When the rapid crossover from the Haldane state to the doubled Mott insulators occurs, the energy gap becomes small and the amplitudes $|z_1|=|z_2|$ of the twist operators decrease. 
This indicates the increase in the localization length of the many-body wavefunction in this regime---while $|z_\alpha|$ converges to unity in the thermodynamic limit in 1D gapped systems, its value can be suppressed when the system size is smaller than or comparable to the localization length.\cite{RestaSorella,NakamuraVoit,NakamuraTodo} 
This behavior can be explained by the suppression of on-site particle number fluctuations when decreasing $L_y/\ell$:  
If such fluctuations are completely absent, the pumping cycle is described by the spin-$1$ chain model \eqref{H_XXZ}, in which the Haldane phase and the doubled Mott insulators cannot be connected without closing a gap. 

\section{Summary\label{summary}}

In summary, we have constructed strongly interacting models of topological pumping by taking the thin-torus limit of 2D QH states. The thin-torus limit of the FQH states is given by CDW ground states; adiabatically connecting between degenerate CDW ground states gives the fractional Thouless pumping. As a more nontrivial example, we have constructed topological pumping which corresponds to the BIQH effect of two-component bosons. The quasi-1D counterpart of the BIQH state is identified as the Haldane phase, and adiabatically connecting between the topological Haldane phase and the trivial doubled Mott insulators constitutes the off-diagonal topological pumping. We have elucidated the nature of the topological pumping via the change in the polarizations between inversion-symmetry-protected quantized values. 
Since the idea of connecting between the Haldane and trivial phases by inversion-symmetry-breaking perturbations does not depend on the details of the system, the obtained off-diagonal topological pumping should not be limited to the thin-torus model considered in this paper. 
While the time-reversal symmetry is broken in QH states and related topological pumping, it is an intriguing direction to construct a bosonic version of time-reversal-symmetric $\mathbb{Z}_2$ pumping\cite{FuKane} which may correspond to 2D bosonic topological insulators.\cite{LiuWen, LiuGuWen}


\begin{acknowledgments}
We are grateful to Masaaki Nakamura, Ippei Danshita, Masahito Ueda, and Norio Kawakami for helpful discussions. M.\ N.\ acknowledges a short-stay research program supported by a Grant-in-Aid for Scientific Research on Innovative Areas ``Topological Materials Science,'' in which this work was initiated. M.\ N.\ was supported by JSPS KAKENHI Grant No.\ JP14J01328 and a JSPS Research Fellowship for Young Scientists. S.\ F.\ was supported by JSPS KAKENHI Grant No.\ JP25800225 and Matsuo Foundation. 
\end{acknowledgments}

\bibliography{BIQHpump_ref.bib}

\end{document}